\documentclass[useAMS,usenatbib]{mn2e}

\title[Near-infrared studies of nova V5579 Sgr]{Nova V5579 Sgr 2008:  Near-infrared studies during  maximum and the early decline phase}

\author[Ashish Raj, N.M. Ashok and D.P.K. Banerjee]{Ashish Raj,$^{}$ N.M. Ashok$^{}$
and D.P.K. Banerjee$^{}$
\thanks{E-mail: ashishr@prl.res.in (AR); ashok@prl.res.in (NMA); orion@prl.res.in (DPKB);} \\
$^{}$Physical Research Laboratory, Navrangpura, Ahmedabad 380009, India\\}

\usepackage[dvips]{graphicx}

\begin{document}

\date{Accepted  Received }

\pagerange{\pageref{firstpage}--\pageref{lastpage}} \pubyear{2010}

\maketitle

\label{firstpage}

\begin{abstract}
We present near-infrared spectroscopic and photometric observations of the nova V5579 Sgr during the
maximum and early decline phase.  The spectra follow the evolution of the nova from peak brightness when the lines
had strong P Cygni profiles to a phase dominated by prominent emission lines.  The spectra during the emission phase
are dominated
by strong  H {\sc i} lines from the Brackett and Paschen series, O\,{\sc i} and
C\,{\sc i} lines. The spectra in the final stages of our observations show a rising continuum towards
longer wavelengths indicating  dust formation. Dust formation in V5579 Sgr is consistent with the presence
of lines of elements with  low
ionization potentials  like Na and Mg in the early spectra. The early presence of such lines had been earlier suggested by us
to be potential indicators of dust formation later in the nova's development.  We also discuss the possibility of  using P Cygni profiles to probe the properties of the erupting white dwarf during the early outburst.

\end{abstract}


\section{Introduction}
Nova V5579 Sgr was discovered on 2008 April 18.784 UT by Nishiyama and Kabashima (Nakano, Nishiyama \& Kabashima 2008) at $V$ = 8.4. Munari et al. (2008) reported a rapid and steady brightening of about 0.7 mag per day in the
initial stages leading to the  possibility of V5579 Sgr reaching naked eye visibility if the trend
continued. However,  the brightening lasted only for 5 days and V5579 Sgr reached a maximum brightness
of $V_{max}$ = 6.65 on 2008 April 23.541 UT as seen from its light curve (Fig. 1).  The early optical spectrum taken during the pre-maximum phase by Fujii (2008) on
2008 April 19.82 UT showed hydrogen Balmer series absorption lines with H$\alpha$ having a prominent P Cygni
profile and also several additional broad absorption lines indicating that  V5579 Sgr is a classical nova.
The infrared spectra taken by Russell et al. (2008) on 2008 May 9 showed lines of O\,{\sc i}, N\,{\sc i},
Ca\,{\sc ii}  and exceptionally strong lines of C\,{\sc i}. The full width at half maximum (FWHM) of the lines
are approximately 1600 km s$^{-1}$. Even though the Fe\,{\sc ii} features were weak Russell et al. (2008) classify
V5579 Sgr to be a Fe\,{\sc ii} type nova. The lines of neutral helium had not yet formed and the strongest
lines  were the O\,{\sc i} lines that are fluorescently excited by Ly$\beta$. The infrared continuum showed
strong thermal emission from dust at a temperature of 1370 K. Subsequent infrared observations
extending to 13.5 ${\rm{\mu}}$m by Rudy et al.(2008) showed significant spectral changes like substantial
decrease in the line strengths and pronounced absorption dip at the line centers. There was an increase in
the dust emission and associated cooling of the dust temperature to 1080 K. The formation of dust can also be seen from the light curve. After reaching the maximum,
V5579 Sgr followed a smooth and fast decline.  This fading was interrupted, about 20 days after discovery,
by a sharp decline seen in the AAVSO light curve consistent with  the formation of dust in the nova ejecta as
reported by Russell et al. (2008). A search in the Digitized Sky Survey (DSS) red image and U.K. Schmidt red
plate by Dvorak (2008) did not reveal any object at the position of V5579 Sgr. With the limiting magnitude of
these surveys being close to 20 magnitudes, V5579 Sgr is one of the large amplitude ($\bigtriangleup$V = 13
magnitudes) novae observed in recent years.

   We present here near-infrared spectroscopic and photometric results of V5579 Sgr based on observations
between 5 and 26 days after the discovery.

\section{Observations}
Near-IR observations were obtained using the 1.2m telescope of Mt.Abu Infrared Observatory from 2008 April 23
to 2008 May 15 . The log of the spectroscopic and photometric observations are given in Table 1 and Table 2
respectively. The spectra were obtained at a resolution of $\sim$ 1000 using a Near-Infrared Imager/Spectrometer
with a 256$\times$256 HgCdTe NICMOS3 array. In each of the $JHK$ bands a set of spectra was taken with the nova
off-set to two different positions along the slit ( slit width 1 arc second). Spectral calibration was done using
the OH sky lines that register with the stellar spectra. The spectra of the comparison star SAO 185320 were taken
at similar airmass as that of V5579 Sgr to ensure that the ratioing process (nova spectrum divided by the standard
star spectrum) removes the telluric features reliably. To avoid artificially generated emission lines in the
ratioed spectrum, the  H\,{\sc i} absorption lines in the spectra of standard star were removed by interpolation
before ratioing. The ratioed spectra were then multiplied by a blackbody curve corresponding to the standard star's
effective temperature to yield the final spectra.

Photometry in the $JHK$ bands was done  in clear sky conditions using the NICMOS3 array in the imaging mode.
Several frames,  in 4 dithered positions, offset by $\sim$ 30 arcsec were obtained in all the bands. The sky
frames, which are subtracted from the nova frames, were generated by median combining the dithered frames.
The star SAO 185779 located close to the nova was used for photometric calibration; the typical errors in the observed magnitudes are
$\pm$0.03. The data is reduced and analyzed using the $IRAF$ package.

\begin{table}
\caption{A log of the spectroscopic observations of V5579 Sgr. The date of
outburst has been assumed to be its detection date viz. 2008 Apr 18.784 UT}
\begin{tabular}{llccc}
\hline
Date & Days      &      & Integration time &     \\
2008 & since     &      &     (s)         &     \\
(UT) & Outburst  & $J$  &    $H$      & $K$ \\

\hline

Apr 23.949  & 5.165   & 20 & 40  & 40\\

Apr 26.966  & 8.182   & 40 & 40 & 20 \\

May 3.947   & 15.163  & 30 & 30 & 60 \\

May 4.977   & 16.193  & 30 & 30 & 60 \\

May 8.972   & 20.188  & 45 & 45 & 60 \\

May 13.914  & 25.13   & 200 & 100 &  \\

\hline
\end{tabular}
\end{table}

\begin{table}
\caption[]{A log of the $JHK$ photometric observations of V5579 Sgr. The date of
outburst has been assumed to be its detection date viz. 2008 Apr 18.784 UT }
\begin{tabular}{llccc}
\hline
Date &   Days   &        & Magnitudes &     \\
2008 &  since   &        &            &     \\
(UT) & outburst &  $J$  & $H$        & $K$  \\
\hline

Apr 23.988 & 5.204  & 4.58     & 4.47  & 4.16  \\
May  1.894 & 13.11  & 5.58     & 5.14  & 4.90  \\
May  8.926 & 20.142 & 6.19     & 5.96  & 4.69  \\
May 14.919 & 26.135 & 6.85     & 5.47  & 4.09  \\
\hline
\end{tabular}
\end{table}

\section{Results}
Before presenting the results proper, we estimate some useful parameters for V5579 Sgr.

\subsection{The pre-maximum rise, outburst luminosity, reddening and distance}
The light curves based on the $V$ band data of AAVSO and $JHK$ magnitudes from Mt. Abu are presented in Fig. 1. There is a good photometric
coverage of the nova's rise to maximum which  lasts for almost 5 days culminating in a peak brightness of
$V_{max}$ = 6.65. From a least square regression fit to the post maximum light curve we estimate $t_2$ to be 8 $\pm$ 0.5 d, making V5579 Sgr one of the fast Fe\,{\sc ii} class of novae in recent years.
 As mentioned earlier V5579 Sgr is also one of the large amplitude novae observed in recent
years with $\bigtriangleup$$V$ = 13 magnitudes. These observed values of
the amplitude and $t_2$  for V5579 Sgr are consistent with its location in  the amplitude versus decline rate plot for
classical novae presented by Warner (2008).
Using the maximum magnitude versus rate of decline (MMRD) relation of della Valle \& Livio (1995), we determine the
absolute magnitude of the nova to be $M_V = -8.8.$ The reddening is derived using the intrinsic colors of
novae at peak brightness, namely $(B-V)$ = 0.23 $\pm$ 0.06, as derived by van den Bergh $\&$ Younger (1987). We have used the optical photometry data
from the  AAVSO to calculate $E(B-V)$. The observed $(B-V)$ = 0.95 $\pm$ 0.06 results in $E(B-V)$ = 0.72 $\pm$ 0.06 and $A_V$ = 2.23 $\pm$ 0.08 for
$R$ = 3.1. Russell et al (2008) estimate $E(B-V)$ = 1.2 using the O\,{\sc i} lines in the spectra obtained on 2008
May 9 but remark that part of the reddening may be local to the nova as dust had already formed. Our
observations, discussed in a later subsection, also clearly show the dust formation in V5579 Sgr.  In their
study of the spatial distribution of the interstellar extinction, Neckel $\&$ Klare (1980) have shown that close to the
direction of V5579 Sgr, $A_V$ steadily increases to a value of $\sim$ 1.8 mag around 2 kpc and flattens after that.
The moderate value of $A_V$ estimated by us appears reasonable even though the nova is located close to the direction
of Galactic Center. Based on the above we obtain a value of the distance $d = 4.4$ $\pm$ 0.2 kpc to the nova.

  \begin{figure}
\includegraphics[bb = 88 240 377 531, width=3.2in,height=3.2in,clip]{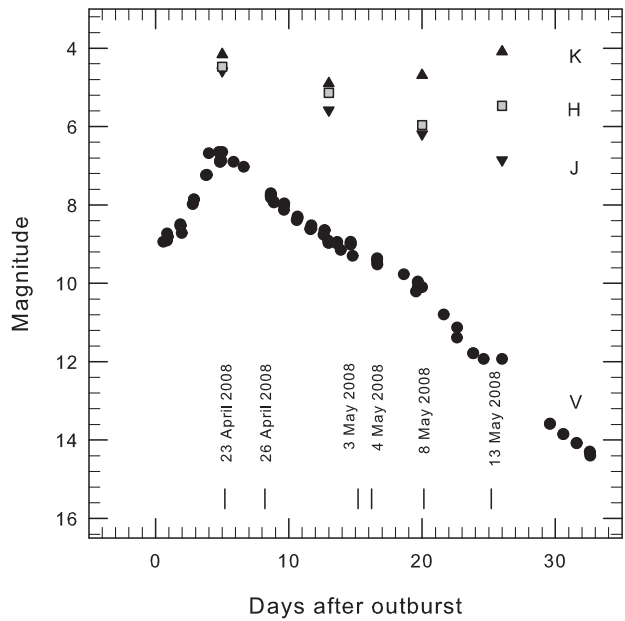}
\caption[]{ The $V$ band lightcurve of V5579 Sgr from AAVSO data. The days of spectroscopic
  observations are indicated by dashes below. The Mt Abu $JHK$ photometric data is also shown.}
 \label{fig1}
 \end{figure}


 \begin{figure}
 \includegraphics[width=3.2in,height=7.2 in,clip ]{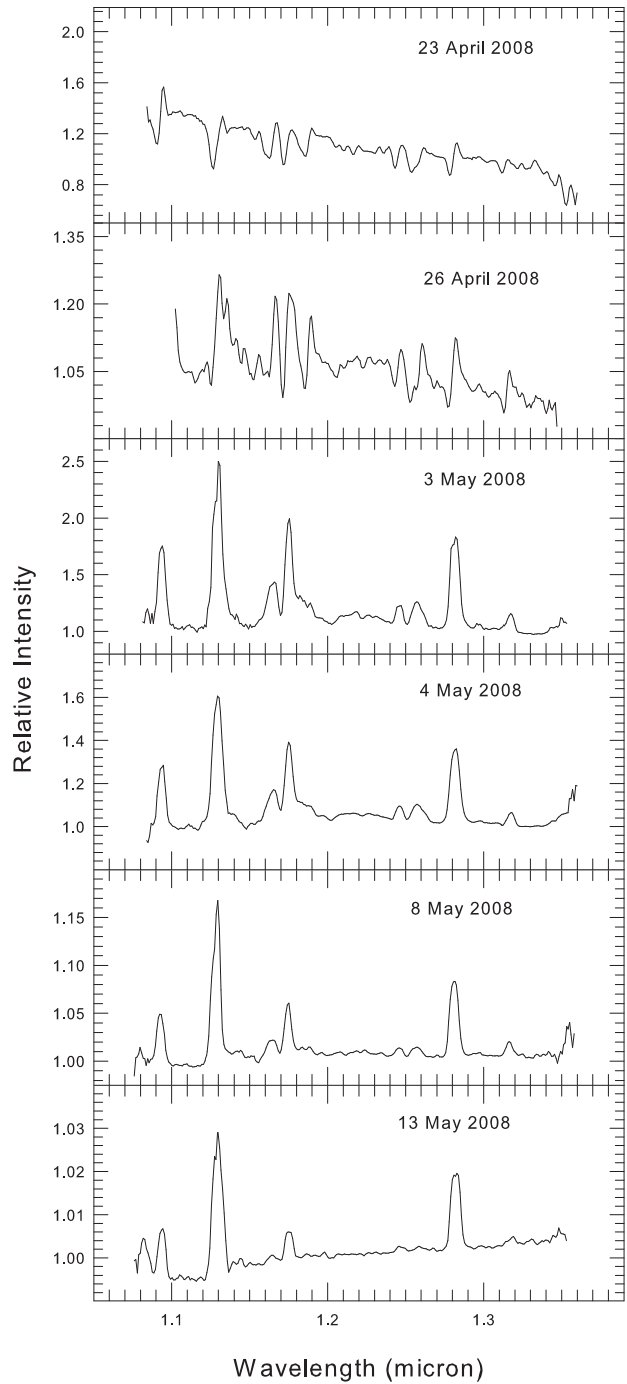}
  \caption[]{ The $J$ band spectra of V5579 Sgr are shown at different epochs. The relative intensity is normalized to unity at 1.25 $\mu m$}
 \label{fig2}
 \end{figure}


\subsection{Line identification, evolution and general characteristics of the $JHK$ spectra}
 The $JHK$ spectra are presented in Figs  2 to 4 respectively; line identification
in graphical and tabular form are shown in Fig. 5 and Table 3 respectively.  The infrared observations presented here cover all the phases with the first infrared spectra taken on 2008 April 23 very close to the visual maximum. These spectra are dominated by  lines of hydrogen, neutral nitrogen and carbon and display deep P Cygni profiles. The emission components of all these lines have become stronger in the spectra taken on 2008 April 26 and by 2008 May 3 all the lines are seen in emission. The typical FWHM of the H\,{\sc i} lines are
1500 km s$^{-1}$. A noticeable feature of these early spectra is the  presence of lines due to Na\,{\sc i} and Mg\,{\sc i}. In the spectra taken on 2008 May 3 the Na\,{\sc i} lines at 1.1381 ${\rm{\mu}}$m, 1.1404 ${\rm{\mu}}$m, 2.1452 ${\rm{\mu}}$m,
2.2056 ${\rm{\mu}}$m and 2.2084 ${\rm{\mu}}$m and  Mg\,{\sc i} lines at 1.1828 ${\rm{\mu}}$m, 1.5040 ${\rm{\mu}}$m, 1.5749 ${\rm{\mu}}$m and 1.7109 ${\rm{\mu}}$m are clearly seen. In an earlier study of V1280 Sco, Das et al. (2008) had suggested
that the presence of spectral lines of low ionization species like Na\,{\sc i} and Mg\,{\sc i} in the early spectra are indicators of low temperature zones conducive to dust formation in the nova ejecta and this is very well borne out in the case of V5579 Sgr.  We would like to point out the presence of a large number of strong lines of neutral carbon. These are typical of Fe\,{\sc ii} type nova as seen in the case of V1280 Sco (Das et al. 2008) and V2615 Oph (Das et al. 2009). The rising continuum is seen in the spectra taken on 2008 May 8 indicating formation of significant amount of dust in the nova ejecta. The dust continuum has started dominating on 2008 May 13.

From the K band spectra  we do not find CO emission bands in the first overtone. However it is possible that such emission may be weakly present but below detection levels. It is thus useful to try and set an upper limit on the strength of the CO emission and hence on the CO mass. An upper limit can be set   by computing a model spectrum for the CO emission and using the criterion that CO emission should be discernible if the calculated model strengths of the CO bands are  atleast  3$\sigma$ times the value of the continuum noise in the CO band region (2.29 - 2.4 ${\rm{\mu}}$m). The model CO spectrum has been computed along the same lines as done for the nova V2615 Oph, where CO was strongly detected, and which is described in details in Das et al. (2009).  The model calculations were done for the temperature range of 2500 K to 4200 K corresponding to the observed values in case of novae where CO has been detected and modeled (first overtone detections have been made  in  V2274 Cyg - Rudy et al. 2003; NQ Vul - Ferland et al. 1979;  V842 Cen - Hyland $\&$ Mcgregor 1989, Wichmann et al. 1990, 1991;  V705 Cas - Evans et al. 1996;  V1419
Aql - Lynch et al. 1995; V2615 Oph - Das et al. 2009; V496 Sct  - Rudy et al. 2009, Raj et al. 2009).
A value of 3 - 5 $\times$ 10$^{-9}$ M$_\odot$ is obtained for the upper limit of M$_{CO}$ using a distance of 4.4 kpc to the nova.

Theoretically, the detailed modeling of Pontefract and Rawlings (2004) for molecule formation and destruction in nova winds, predicts that CO should form early after the outburst, remain constant in strength for $\sim$ 15 days thereafter and then get rapidly destroyed. Fairly consistent with this picture, most of the CO detections outlined above have indeed been reported early after the outburst (see Das et al. 2009 for a detailed discussion). Thus, in the present case too, CO emission could have been expected. Its absence indicates that  it either is   present
 but below detection levels or that it did not form for  reasons  which are not clearly understood.

 \begin{figure}
 \includegraphics[bb= 38 169 288 720, width=3.1in,height=7.2in, clip]{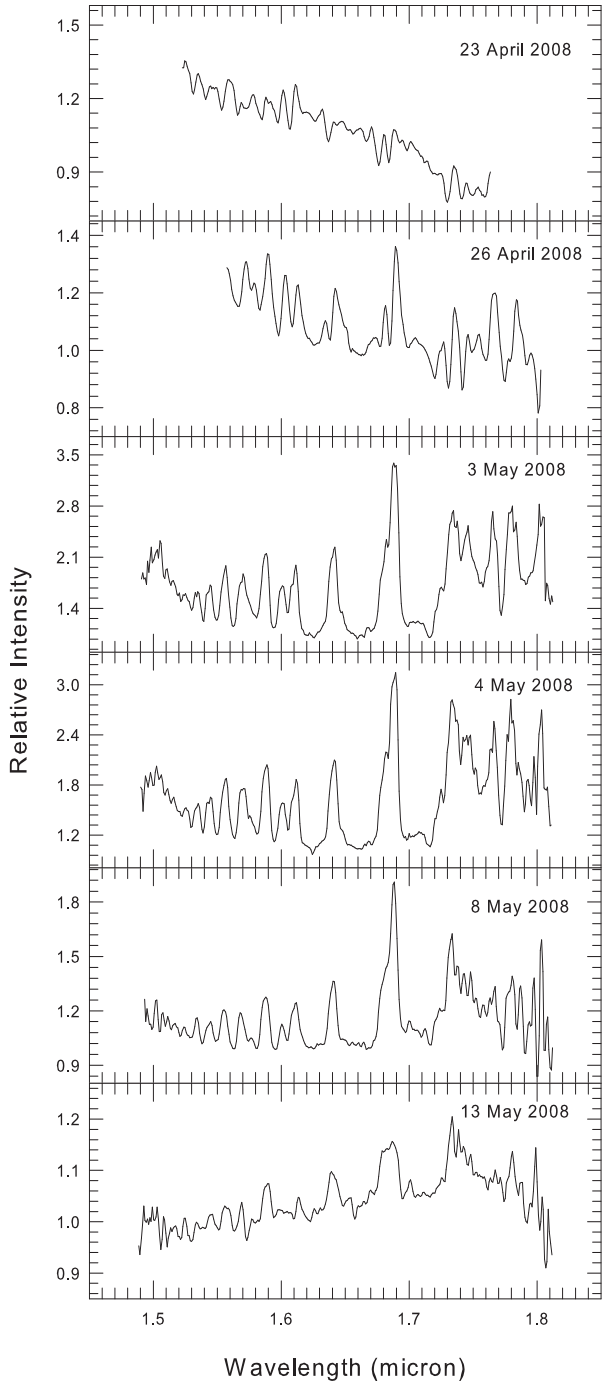}
  \caption[]{The $H$  band spectra of V5579 Sgr are shown at different epochs. The
   relative intensity is normalized to unity at 1.65 $\mu m$ }
  \label{fig3}
  \end{figure}


  \begin{figure}
  \includegraphics[width=3.2in,height=6.0in, clip]{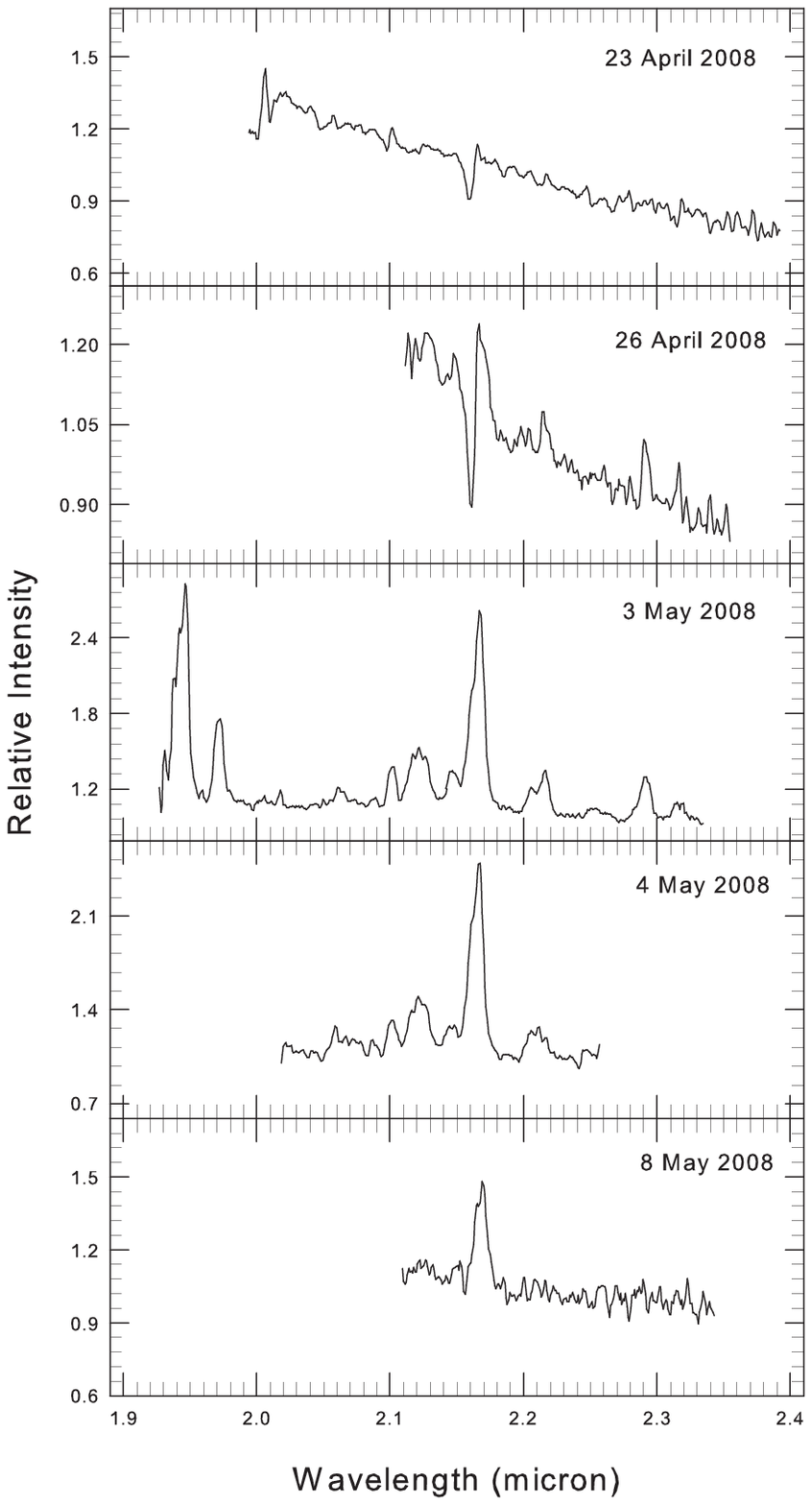}
 \caption[]{The $K$  band spectra of V5579 Sgr are shown at different epochs. The
   relative intensity is normalized to unity at 2.2 $\mu m$}
     \label{fig4}
  \end{figure}


  \begin{figure}
\includegraphics[bb= 45 41 451 757, width=3.2in,height=7.0in, clip]{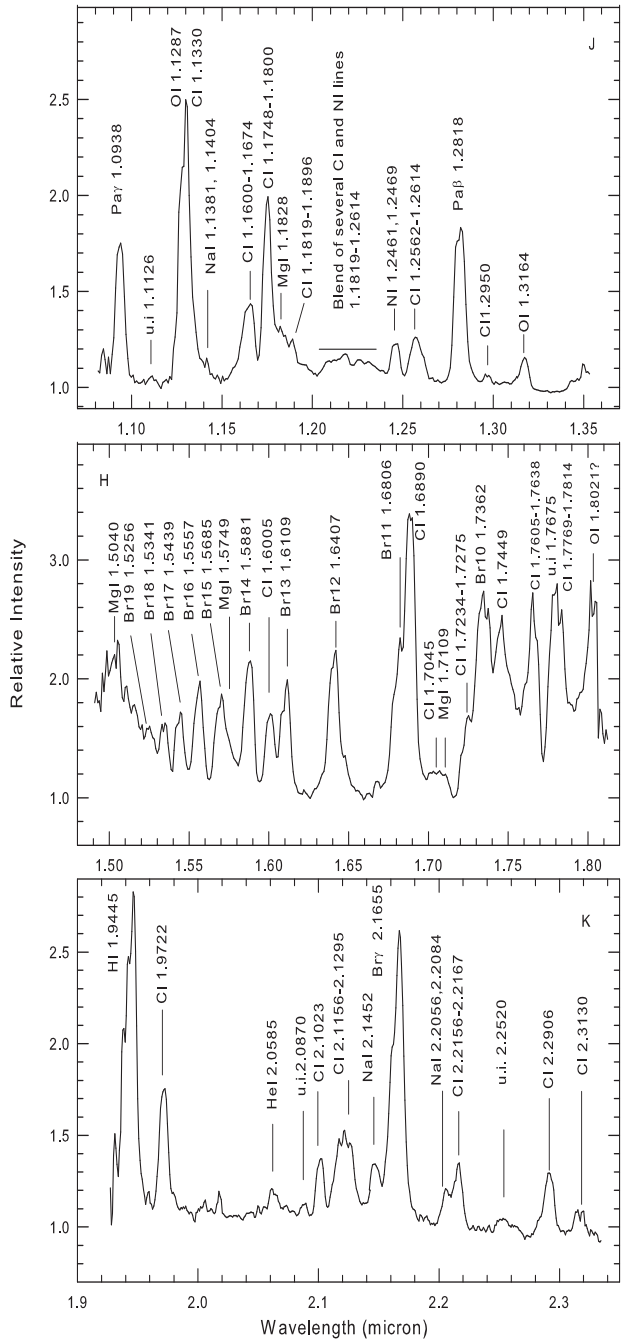}
  \caption[]{ The line identification is shown for the spectra of 2008 May 3 in the
   $JHK$ bands as given in Table 3.}
  \label{fig5}
  \end{figure}

\subsection{Fireball phase}
As noted earlier, V5579 Sgr showed pre-maximum brightening. After its discovery on
2008 April 18.784 UT by Nishiyama \& Kabashima (Nakano 2008) at 8.4 mag, it brightened by nearly 1.8 mag over next five days to reach maximum of $V$ = 6.65 mag on 2008 April 23.541 UT. This pre-maximum rise is well observed at optical wavebands. Our first near-IR photometric observations are available for 2008 April 23.988 UT close to the optical maximum. We have studied the fireball phase by obtaining spectral energy distribution (SED) covering the optical and near-IR region. The following optical magnitudes $B$=7.6, $V$=6.87, $R_C$=6.23 and
$I_C$=5.55 at maximum brightness from AAVSO along with the present $JHK$ magnitudes of 2008 April 23.988 UT were used in deriving the SED. The observed magnitudes were corrected for extinction using Koornneef's (1983) relations and $A_V$=2.23 mag. From a blackbody fit to the SED shown in top panel of Fig. 6, we obtain a formal  temperature of $T_{bb}$ = 8900 $\pm$ 400 K which is consistent with the A to F spectral type for the pseudo-photospheres displayed by novae at outburst (Gehrz 1988). However, this temperature $T_{bb}$ estimate is likely to have additional errors as all the observed flux values used in the fit lie on the Rayleigh-Jeans part of the spectral energy distribution. Further, the $B$ band flux value is also susceptible to significant errors since it has the largest interstellar extinction correction. The blackbody angular diameter $\theta_{bb}$ in arcseconds is calculated using the relation given by Ney $\&$ Hatfield (1978), namely,

\begin{equation}
\nonumber
\theta_{bb} = 2.0 \times 10{^{\rm 11}} (\lambda F_{\lambda})_{max}^{1/2}\times T_{bb}^{-2}
\end{equation}
where $ (\lambda F_{\lambda})_{max}$ is in W cm$^{-2}$  and $T_{bb}$ is in Kelvin. From the model blackbody fit of 8900 K shown in Fig. 6 (upper panel), we find $(\lambda F_{\lambda})_{max}$ = 3.64 $\times$ 10${^{\rm -14}}$  W cm$^{-2}$. We accordingly obtain a value of $\sim$ 0.5 milliarcsec for the angular diameter. This value for the angular diameter can be used to estimate the distance to the nova by invoking constant expansion rate for the ejecta and
the relation given by Gehrz (2008), namely,

\begin{equation}\nonumber
$ $ d = 1.15 \times 10{^{\rm -3}} (V_{ej})t/ \theta_{bb} 
\end{equation}

where $d$ is in kpc, $V_{ej}$ in km s$^{-1}$, $t$ in days since outburst began and $\theta_{bb}$ in milliarcsec. Taking a value of 1500 km s$^{-1}$, the typical FWHM of H\,{\sc i} lines for $V_{ej}$ and $t = 5$ d corresponding to the epoch of optical maximum we get d = 17.3 kpc which is four times larger than the estimate done in section 3.1 using the MMRD relation.
The reason for this discrepancy is not clear. A likely reason is that the  pseudo photosphere behaves as a grey body with reduced emissivity in the fireball phase. The estimate of $\theta_{bb}$ will always be a lower limit since it is applicable for a blackbody (Ney $\&$ Hatfield 1978; Gehrz et al. 1980). For a grey body, the observed angular size can be larger, since the right hand side of  equation 1 should be divided by $\epsilon^{1/2}$, where $\epsilon$ the emissivity has a value less than unity. A similar discrepancy was noticed by Das et al. (2008) in the case of V1280 Sco, where the blackbody angular diameters were smaller by a factor of three than the values derived from the interferometric measurements. It should be noted that the value of $t$ is likely to have an error due to the uncertainty in the determination of the start of the outburst. For V5579 Sgr, Yamaoka (2008) and Liller (2008) report that no object brighter than 11.5 and 11.0 mag respectively was seen on their petrol images taken around 2008 April 15.743 UT and 2008 April 16.22 UT. Thus our estimate of $t$ is likely to have an error of $\sim$ 2 days which worsens slightly more the discrepancy between the distances determined using the MMRD and the blackbody angular diameter relations.

\begin{table}
\caption[]{A list of the lines identified from the $JHK$ spectra shown in Figure 5. The additional lines contributing to the identified lines are listed and the unidentified lines are
     mentioned as u.i.}
\begin{tabular}{llllll}
\hline\\
Wavelength & Species  & Other contributing  \\
(${\rm{\mu}}$m) & &lines and remarks   \\
\hline
\hline \\

1.0938   & Pa $\gamma$        &      \\
1.1126   & u.i                & Fe\,{\sc ii} ? \\
1.1287   & O\,{\sc i}         &      \\
1.1330   & C\,{\sc i}         &        \\
1.1381   & Na\,{\sc i}        &    C\,{\sc i} 1.1373  \\
1.1404   & Na\,{\sc i}        &    C\,{\sc i} 1.1415  \\
1.1600-1.1674   & C\,{\sc i}  & strongest lines at 1.1653,\\
                &             &           1.1659,1.16696    \\
1.1746-1.1800   & C\,{\sc i}  & strongest lines at 1.1748,  \\
                &             &          1.1753,1.1755      \\
1.1828          & Mg\,{\sc i} &               \\
1.1819-1.2614   & several C\,{\sc i}  & strongest lines at 1.1880,  \\
                & and N\,{\sc i}      &           1.1896 \\
1.2461,1.2469 & N\,{\sc i} & blended with O\,{\sc i} 1.2464 \\
1.2562,1.2569 & C\,{\sc i} & blended with O\,{\sc i} 1. 2570 \\
1.2818   & Pa $\beta$         &     \\
1.2950   & C\,{\sc i}         &     \\
1.3164   & O\,{\sc i}         &     \\
1.5040   & Mg\,{\sc i}        &  blended with Mg\,{\sc i} 1.5025,\\
         &                    &         1.5048 \\
1.5256   & Br 19              &           \\
1.5341   & Br 18              &           \\
1.5439   & Br 17              &           \\
1.5557   & Br 16              &           \\
1.5701   & Br 15              &           \\
1.5749   & Mg\,{\sc i}        & blended with Mg\,{\sc i} 1.5741,  \\
         &                    &           1.5766,C\,{\sc i} 1.5788 \\
1.5881   & Br 14              &  blended with C\,{\sc i} 1.5853    \\
1.6005   & C\,{\sc i}         &           \\
1.6109   & Br 13              &           \\
1.6407   & Br 12              &           \\
1.6806   & Br 11              &           \\
1.6890   & C\,{\sc i}         &           \\
1.7045   & C\,{\sc i}         &           \\
1.7109   & Mg\,{\sc i}        &               \\
1.7234-1.7275 & C\,{\sc i}    & several C\,{\sc i} lines  \\
1.7362   & Br 10              &  affected by C\,{\sc i} 1.7339 line    \\
1.7449 & C\,{\sc i}           &           \\
1.7605-1.7638 & C\,{\sc i}    &           \\
1.7675  & u.i                 &           \\
1.7769-1.7814 & C\,{\sc i}    &           \\
1.8021  & O\,{\sc i} ?        &           \\
1.9445  & H\,{\sc i}          &           \\
1.9722  & C\,{\sc i}          &           \\
2.0585 & He\,{\sc i}          &           \\
2.0870  & u.i                 &           \\
2.1023 & C\,{\sc i}           &           \\
2.1156-2.1295 & C\,{\sc i}    &           \\
2.1452 & Na\,{\sc i}          &           \\
2.1655   & Br $\gamma$        &           \\
2.2056,2.2084 & Na\,{\sc i}   &           \\
2.2156-2.2167 & C\,{\sc i}    &           \\
2.2520  & u.i     &    \\
2.2906 & C\,{\sc i}           &           \\
2.3130 & C\,{\sc i}           &           \\

\hline
\end{tabular}
\end{table}

 \begin{figure}
\includegraphics[bb=29 518 288 720,width=3.2in,height=3.2in,clip]{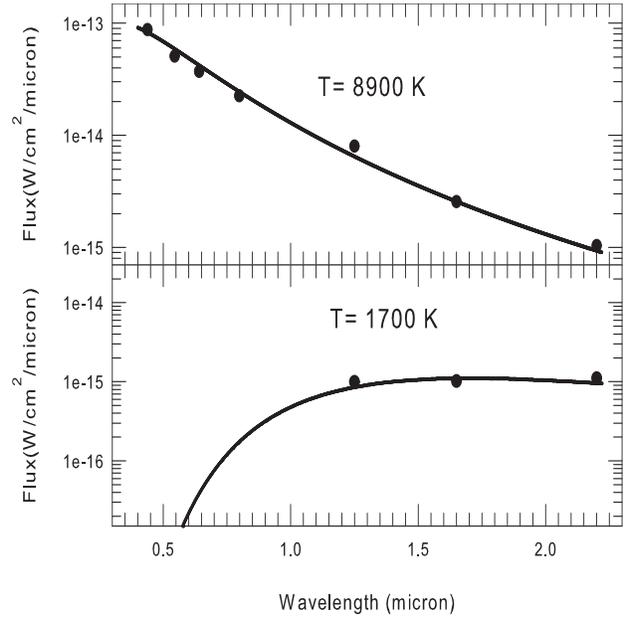}
\caption[]{ The top panel shows spectral energy distribution for the fireball phase data of April 23.988 UT with blackbody temperature fit of 8900 K. The bottom panel shows a similar fit for the data of May 14.919 UT with blackbody temperature fit of 1700 K after dust formation.}
  \label{fig8}
  \end{figure}

\subsection{Dust formation and ejecta mass estimate}
The light curve showed a sharp fall about 15 days after the visual maximum indicating onset of dust formation. The thermal emission from the dust contributes to the near-IR bands and one expects a brightening at these wavelengths. The present near-IR photometric observations presented in Fig. 1 clearly show the onset of dust formation associated with the fall in the visual light curve around 2008 May 8 accompanied simultaneously  thereafter by a steady increase in the near-IR magnitudes, especially in the $K$ band.
The SED of the dust component in the ejecta is constructed using the observed $JHK$ magnitudes on 2008 May 14.919 UT and shown in the lower panel Fig. 6.  The contribution of the thermal emission from the dust is seen increasing upto the $K$  band  indicating that it may peak at even longer wavelengths.
We estimate a value of 1700 $\pm$ 200 K for the temperature of the dust shell. However, this estimate of temperature for the dust shell has large uncertainty as  observations in only three wavelengths are used in fitting the SED of which the dust is contributing mostly in the K band. Russell et al (2008) have estimated the dust shell temperature to be 1370 K based on the observations spanning the wavelength range upto 5.2 ${\rm{\mu}}$m. The likely reason for the higher value for the dust temperature derived by us is the restricted spectral coverage extending upto K band only that may have emphasized the contribution at shorter wavelengths.

 \begin{figure}
\includegraphics[bb= 86 249 388 541, width=3in,height=3.0in, clip]{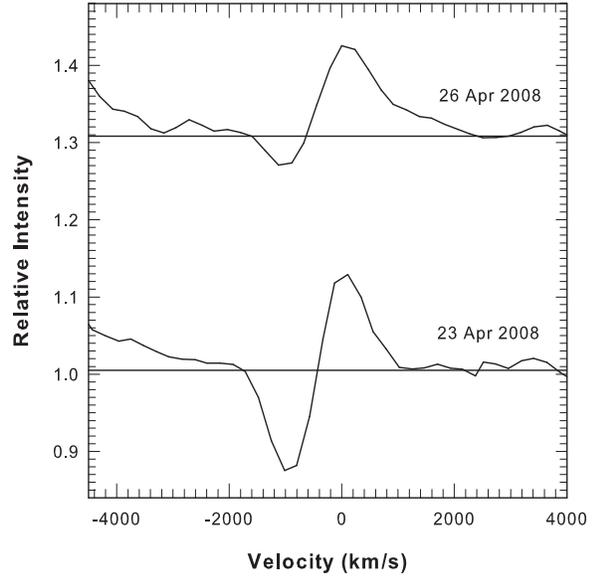}
  \caption[]{ The time evolution of $Pa\beta$ 1.2818 $\mu$m line showing the evolution of the P Cygni feature. For more details see section 3.5.}
  \label{fig7}
  \end{figure}

The mass of the dust shell can be calculated from the thermal component of the SED of 2008 May 14 UT shown in Fig. 6. Woodward et al. (1993) have given the expression for the mass of the dust shell as
$M_{dust}$ = 1.1 $\times$ 10${^{\rm 6}}$ $(\lambda F_{\lambda})_{max}$ $d^2$ / $T_{dust}^6$ .
In the expression above, mass of the dust shell $M_{dust}$  is in units of $M_\odot $ , $(\lambda F_{\lambda})_{max}$ is in W cm$^{-2}$  measured
at peak of the SED, the black-body temperature of the dust shell $T_{dust}$ is in units of 10${^{\rm 3}}$ K, and the distance to the nova $d$ is in kpc. It is assumed that the dust is composed of carbon particles of size less than 1 ${\rm{\mu}}$m with a density of 2.25 gm cm$^{-3}$. The early dust formation, high dust temperature and SED in the near-infrared that resembles black-body in the case of V5579 Sgr are indicative of presence of carbon grains in the dust shell (Clayton $\&$ Wickramsinghe 1976). In addition the occurrence of strong carbon spectral features in the observed spectra indicate that our assumption that the dust is made up of carbon/graphite
is reasonable. We obtain $M_{dust}$ = 2.12 $\times$ 10${^{\rm -9}}$  $M_\odot $  for 2008 May 14 taking the observed parameters of
$(\lambda F_{\lambda})_{max}$ = 2.42 $\times$ 10${^{\rm -15}}$  W cm$^{-2}$, $T_{dust}$= 1.7 $\times$ 10${^{\rm 3}}$
 K and $d$ = 4.4 kpc. The observed ratio for the $M_{gas}$ to $M_{dust}$ range from 1.8 $\times$ 10${^{\rm 2}}$  for LW Ser that formed optically thick dust
shell (Gehrz et al. 1980) to 2.5 $\times$ 10${^{\rm 4}}$  for V1425 Aql that formed optically thin dust shell (Mason et al. 1996). Taking a canonical value of 200 for the gas to dust ratio, we get $\sim$ 4.2$\times$ 10${^{\rm -7}}$  $M_\odot $ for the gaseous component of the ejecta. This value is smaller than the typically observed value
of 10${^{\rm -4}}$  to 10${^{\rm -6}}$ $M_\odot$ in novae. One definite reason for the dust mass being underestimated is that our SED is based on data extending upto only 2.2 $\mu$m and certainly neglects contribution from dust emission at longer wavelengths. In this process we are overestimating the derived dust
temperature considerably as is evident from Russell et al. (2008) who get $T_{dust}$ = 1370 K on 2008 May 9, five days before the date being considered in this analysis,
which further cools down quickly to 1070 K by 2008 May 22 (Rudy et al. 2008). The formulation used for the dust mass estimate is very sensitive to $T_{dust}$. Further, since the Rudy et al. (2008) and Russell et al. (2008) reports show that the dust emission peaks at longer wavelengths, we are also underestimating $(\lambda F_{\lambda})_{max}$. Correct use of  both $T_{dust}$ and $(\lambda F_{\lambda})_{max}$ values should considerably enhance the dust mass estimate made here.

We have alternatively explored the possibility of  estimating the ejecta mass using recombination line analysis of H\,{\sc i}. However, we find that the  strengths of these lines, relative to each other, deviate considerably from Case B values on all epochs indicating that the lines are optically thick. Hence we are unable to estimate the ejecta mass from recombination analysis.

\subsection{A discussion of the  P Cygni phase}

We qualitatively discuss the P Cygni profiles seen around maximum light as they can help estimate the radius of the white dwarf's (WD)
photosphere ($r_{ph}$) at this epoch. Kato \& Hachisu (1994) have shown from theoretical considerations of the photospheric optical depth
that $r_{ph}$ $\geq$ 100 $R_{\odot}$ at maximum. This is a substantial increase by a factor of almost $10^4$ in the WD's radius between quiescence to maximum. Subsequent refinements in their modeling of novae light curves reaffirm that  after  the thermonuclear runaway sets in on a mass-accreting WD, its envelope expands greatly to
$r_{ph}$$\geq$  100 $R_{sun}$ (Hachisu \& Kato, 2001;  Hachisu \& Kato, 2006 ) the evolution of $r_{ph}$ with time is illustrated diagrammatically in the above works. It is  desirable to have observational confirmation for such estimates of $r_{ph}$ and  P Cygni profiles  may provide an assessment of this  physical parameter.

The generic formation of P Cygni profiles, following Lamers \& Cassinelli (1999),  can be  understood   by considering
a  spherical symmetric outflowing wind (in our case  the mass loss from the nova outburst) in which the velocity
necessarily increases outwards i.e., the wind is accelerated outwards till it reaches a terminal velocity. To the outside observer, it is the matter in the form of a tube in
front of the stellar disc which scatters light from the continuum of
the star that is responsible for the absorption component of the P
Cygni profile (see Fig. 2.4 in Lamers \& Cassinelli (1999)).
The ratio of the strength of the emission and absorption components of the P Cygni profile depends on the size $r_w$ of the wind region
(i.e., size of the ejected material) relative to the size  of the star. If the star is large compared to the size of the wind
region the emission will  be smaller than the absorption. When the wind region is large compared to the star's size we expect emission
to dominate - this can be seen  geometrically as the volume of the emitting gas becomes much larger than the volume causing the absorption component.
Observationally consistent with this scenario, it is known  that prominent P Cygni profiles in  a nova outburst are inevitably seen
and reported at  maximum light and on one or two  days following it.  At such an epoch, it is therefore reasonable to use an
approximation that the wind region size $r_w$ is of the order of the stellar size $r_{ph}$.  Since the wind region size $r_w$ can be approximated
kinematically ($r_w$ $\sim$ $v_{mean}$$\times$$t$, $v_{mean}$= mean velocity of ejecta, $t$ = time after outburst), a qualitative estimate can be made of $r_{ph}$.
This could be used as the rationale in estimating $r_{ph}$. As the ejected matter (the wind region)  keeps expanding to larger sizes at
 later times following the maximum, the emission component strengthens and finally begins to dominate. This expected behaviour is reasonably in accordance with the early P Cygni profiles seen in V5579 Sgr as shown in Fig. 7.

A simplistic order-of-magnitude estimate for $r_{ph}$ may be obtained in the following way. Let a typical velocity of $v_{mean}$ $\sim$ 1000 km s$^{-1}$
for the nova ejecta be assumed.  Velocities in nova ejecta can range from few hundreds to few thousands of  km s$^{-1}$ and $v_{mean}$ = 1000 km s$^{-1}$
is a  fairly representative value. For our choice of $v_{mean}$, and taking $t$ = 1 d  as the typical timescale when prominent P Cygni profiles
are generally seen, the value of $r_w$ $\sim$ 120 $R_{sun}$  encouragingly matches that expected for $r_{ph}$.
However, we emphasize  that this is purely a qualitative estimate. We hope to undertake  a detailed modelling, which takes into account a realistic wind-velocity law in the ejecta,  to try and   reproduce  observed P Cygni profiles in novae and their evolution with time.

\section{Summary}
We have presented near-infrared spectroscopy and photometry of nova V5579 Sgr which erupted on 2008 April 19. From the optical lightcurve,  the distance to the nova is estimated to be  $4.4$ $\pm$ 0.2 kpc. The infrared spectra indicate that the nova is of the Fe II class.   Evidence is seen from the $JHK$ photometry and spectra for the formation of dust in the nova in mid-May 2008. In this context, the presence of emission lines from low ionization species like Na and Mg in the early spectra and subsequent formation of the dust supports the predictive property of these lines as indicators of dust formation as proposed by Das et al (2008). It may be noted that V5579 Sgr is one of the few fast Fe\,{\sc ii} class of novae ($t_2$ = 8 d) that formed dust. We have indicated the possible usefulness of  the P Cygni profiles seen in the novae spectra around maximum brightness to study the physical parameters related to the central white dwarf.

\section{Acknowledgments}
The research work at Physical Research Laboratory is funded by the Department of Space, Government of India. We thank the AAVSO (American Association of Variable Star Observers) for the use of their optical photometric data. We thank the referee for his helpful comments.

\label{lastpage}
\end{document}